\documentclass[sigconf]{acmart}
\usepackage{enumitem}
\usepackage{multirow}
\usepackage{booktabs}
\usepackage{subcaption} 
\usepackage{hyperref}
\usepackage{colortbl}
\usepackage{graphicx}   
\usepackage{makecell}   
\usepackage{xcolor}     

\usepackage{amssymb}    
\usepackage{pifont}     
\definecolor{linecolor}{gray}{.85} 
\definecolor{linecolor2}{gray}{.9} 
\definecolor{linecolor1}{gray}{.95} 
\AtBeginDocument{%
  }

\setcopyright{acmlicensed}
\copyrightyear{2018}
\acmYear{2018}
\acmDOI{XXXXXXX.XXXXXXX}
\acmConference[Conference acronym 'XX]{Make sure to enter the correct
  conference title from your rights confirmation email}{June 03--05,
  2018}{Woodstock, NY}
\acmISBN{978-1-4503-XXXX-X/2018/06}

\begin{document}

\title{FAIR: Focused Attention Is All You Need for Generative Recommendation}

\author{Longtao Xiao}
\authornote{Both authors contributed equally to this research.}
\affiliation{%
  \institution{Huazhong University of Science and Technology}
  \city{Wuhan}
  \country{China}
}
\email{xiaolongtao@hust.edu.cn}

\author{Haolin Zhang}
\authornotemark[1]
\affiliation{%
  \institution{Shenzhen International Graduate School, Tsinghua University}
  \city{Shenzhen}
  \country{China}
}
\email{zhanghaolin24@mails.tsinghua.edu.cn}

\author{Guohao Cai}
\affiliation{%
  \institution{Huawei Noah’s Ark Lab}
  \city{Shenzhen}
  \country{China}
}
\email{caiguohao1@huawei.com}

\author{Jieming	Zhu}
\affiliation{%
  \institution{Huawei Noah’s Ark Lab}
  \city{Shenzhen}
  \country{China}
}
\email{jiemingzhu@ieee.org}

\author{Yifan Wang}
\affiliation{%
  \institution{Huazhong University of Science and Technology}
  \city{Wuhan}
  \country{China}
}
\email{d202381481@hust.edu.cn}

\author{Heng Chang}
\affiliation{%
  \institution{Huawei Technologies Co., Ltd}
  \city{Shanghai}
  \country{China}
}
\email{changh.heng@gmail.com}

\author{Zhenhua	Dong}
\affiliation{%
  \institution{Huawei Noah’s Ark Lab}
  \city{Shenzhen}
  \country{China}
}
\email{dongzhenhua@huawei.com}

\author{Xiu Li}
\affiliation{%
  \institution{Shenzhen International Graduate School, Tsinghua University}
  \city{Shenzhen}
  \country{China}
}
\email{li.xiu@sz.tsinghua.edu.cn}
\authornote{Corresponding author}

\author{Ruixuan	Li}
\affiliation{%
  \institution{Huazhong University of Science and Technology}
  \city{Wuhan}
  \country{China}
}
\email{rxli@hust.edu.cn}
\authornotemark[2]
\renewcommand{\shortauthors}{Xiao et al.}

\begin{abstract}

Recently, transformer-based generative recommendation has garnered significant attention for user behavior modeling. However, it often requires discretizing items into multi-code representations (e.g., typically four code tokens or more), which sharply increases the length of the original item sequence. This expansion poses challenges to transformer-based models for modeling user behavior sequences with inherent noises, since they tend to overallocate attention to irrelevant or noisy context. To mitigate this issue, we propose \textbf{FAIR}, the first generative recommendation framework with focused attention, which enhances attention scores to relevant context while suppressing those to irrelevant ones. Specifically, we propose (1) a focused attention mechanism integrated into the standard Transformer, which learns two separate sets of Q and K attention weights and computes their difference as the final attention scores to eliminate attention noise while focusing on relevant contexts; (2) a noise-robustness objective, which encourages the model to maintain stable attention patterns under stochastic perturbations, preventing undesirable shifts toward irrelevant context due to noise; and (3) a mutual information maximization objective, which guides the model to identify contexts that are most informative for next-item prediction. We validate the effectiveness of FAIR on four public benchmarks, demonstrating its superior performance compared to existing methods.

\end{abstract}




\maketitle

\begin{figure}[htbp]
    \centering
    \includegraphics[width=\linewidth]{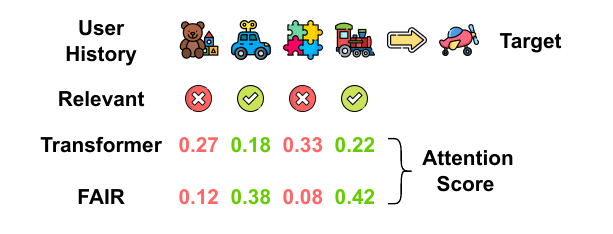}
    \caption{Transformer often tends to overallocate attention to irrelevant context (e.g., attention noise). However, FAIR enhance attention to relevant context while suppressing noise.}
    \label{fig:case study}
\end{figure}

\section{Introduction}
Generative recommendation \cite{rajput2023recommender, zhai2024actions, deng2025onerec, zheng2024adapting} is emerging as a promising paradigm for sequential recommendation tasks \cite{kang2018self, wang2019sequential}. By tokenizing the user's historical interaction behaviors (usually represented by interacted items) into a list of discrete codes (also denoted as semantic IDs \cite{rajput2023recommender}), it is possible to fully leverage the powerful language modeling capability of transformer to autoregressively generate the codes, and then map these codes back to the corresponding items.
While tokenizing each item into four discrete codes is a common practice \cite{rajput2023recommender,zheng2024adapting}, recent studies \cite{hou2025generating, hou2023learning} have demonstrated that using longer codes can enhance the representational capacity of code space and thus improve recommendation performance. However, this also incurs a serious issue, the user's original historical interaction sequence is drastically expanded. As an item sequence inherently contain noisy and irrelevant items, the exapanded sequence can lead to the issue of \textit{attention noise} \cite{chen2023accumulated} in transformer. 
That is, as depicted in Figure \ref{fig:case study}, the Transformer tends to allocate excessive attention to irrelevant items, resulting in insufficient focus on relevant ones, which ultimately harms recommendation performance. For clarity, we define relevant context as the user’s recent interactions identified based on collaborative co-occurrence statistics and semantic relevance, while interactions outside this scope are treated as noise. To obtain an attention score for each item, we compute the average attention across all discrete codes in its code sequence and visualize the aggregated scores. The results demonstrate that the Transformer pays excessive attention to noise, while informative relevant contexts receive inadequate emphasis. As a result, these critical contexts are overshadowed, hindering effective modeling of long code sequences.


In this work, we aim to amplify the transformer’s attention to relevant contexts while mitigating the influence of noise. The core idea is to first identify which parts of the user’s historical interactions are directly relevant for modeling user interests. However, due to the presence of irrelevant and distracting information in the interaction history, it is equally critical to ensure the model exhibits sufficient robustness against such noise. Achieving this is a non-trivial task: unlike retrieval tasks where relevant contexts can be explicitly determined, recommendation tasks require implicit inference to discern which parts of the context are informative for predicting the next interaction. Therefore, to capture user interest from historical code sequence, it is essential to endow the model with the capability to amplify its attention to these contexts while suppressing the impact of noise.



To this end, we propose FAIR, a novel generative recommendation framework which enables to enhance attention to relevant context while suppressing noise for the
first time. Our framework contains three technical aspects: (1) \textbf{Focused Attention Mechanism}. Inspired by the differential amplifier \cite{sackinger1987versatile} in signal processing, which attenuates noise and interference, we design a focused attention mechanism that learns two distinct sets of query and key matrices. The attention scores computed from these two sets are then subtracted to produce the final attention distribution, thereby highlighting informative contexts. (2) \textbf{Noise-Robustness Task}. We simulate noisy inputs by randomly masking or replacing parts of the original sequence and impose strong consistency constraints to align their hidden representations with those of the clean inputs, thus improving the model’s robustness to random perturbations. (3) \textbf{Mutual Information Maximization Task}. We maximize the mutual information between the final-layer hidden representation of the input and its corresponding target representation. This encourages the extraction of contexts that are most informative for next-item prediction, enabling the model to more effectively focus on relevant information.

In summary, our contributions can be concluded as follows:
\begin{itemize}[left=0pt]
    \item To the best of our knowledge, we propose FAIR, the first generative recommendation framework with focused attention, which enhances attention scores to relevant context while suppressing those to irrelevant ones.
    \item We introduce a focused attention mechanism, a noise-robustness task, and a mutual information maximization task to enhance the transformer’s ability to identify relevant contexts and allocate greater attention to them, preventing critical information from being overshadowed by attention noise.
    \item Extensive experiments on four public recommendation benchmarks demonstrate FAIR’s superiority over existing methods, encompassing both generative and traditional models.
\end{itemize} 

\section{Related Work}
Sequential recommendation approaches model user behavior as a chronologically ordered sequence of interactions, aiming to predict the next item a user will be interested in. Based on the reliance on dot-product (cosine) similarity with approximate nearest neighbor (ANN) search \cite{andoni2018approximate}, existing methods can be divided into traditional paradigms and generative paradigms \cite{xiao2025progressive}.

\subsection{Traditional Paradigms}
Early methods \cite{rendle2010factorizing, he2016fusing} primarily apply Markov Chains to calculate transition probabilities between items. However, the Markov assumption limits these methods' ability to model long-range dependencies. With the success of deep learning, recurrent neural network (RNN) is introduced to sequential recommendation. GRU4Rec \cite{jannach2017recurrent} introduces several modifications such as ranking loss to classic RNN structure to make it more suitable for recommendation task. But RNN-based paradigms struggle with long sequence modeling and training efficiency. To solve this issue, SASRec \cite{kang2018self} first leverages decoder-only self-attention model to capture both long-range semantics and sparse actions in interaction sequences. BERT4Rec \cite{sun2019bert4rec} leverages mask language model (MLM) strategy for the modeling of item sequences. Building on the masking technique, S\textsuperscript{3}-Rec \cite{zhou2020s3} optimizes four self-supervised objectives to enhance data representation through the principle of mutual information maximization (MIM). These approaches typically learn a high-dimensional embedding for each item and use ANN search among a set of candidate items to predict the next item.

\subsection{Generative Paradigms}
Unlike traditional paradigms, which rely on dot-product (cosine) similarity and external ANN search systems for top-k recommendation, generative paradigms reframe recommendation as a sequence generation task. Typical generative recommendation methods \cite{rajput2023recommender, deng2025onerec} leverage semantic embeddings from items to produce codes, then use generative models to predict the next codes, finally map these codes to corresponding items. Transformer-based models dominate the generative model landscape. P5 \cite{hua2023index} finetunes pretrained T5 to perform recommendation tasks using natural language prompts. GPTRec \cite{petrov2023generative} uses item titles as semantic representations and introduces pretrained GPT-2 model into generative recommendation. TIGER \cite{rajput2023recommender} pioneers the use of vector quantization for producing compact codes and trains a T5 encoder-decoder model from scratch to autoregressively predict the next codes. HSTU \cite{zhai2024actions} introduces pointwise aggregated attention to capture the intensity of user preferences and be customized for non-stationary vocabularies. However, previous works have not noticed the potential attention noise issue in Transformer-based generative models.

In this paper, we propose a novel framework named \textit{FAIR} for the first time to endow generative models with the ability to enhance attention to relevant context while suppressing noise.


\section{Method}
In this section, we present FAIR, a non-autoregressive (NAR) generative recommendation model designed to identify and amplify attention to relevant contexts while mitigating noise. First, we formulate the task in Section \ref{section Problem Formulation}. Then, we introduce the focused attention mechanism in Section \ref{section Focused Attention}. Next, Section \ref{section Noise-Robustness Task} introduces the noise-robustness task and its optimization objective, while Section \ref{section Mutual Information Maximization Task} presents the mutual information maximization task along with its optimization objective. Finally, we describe the overall training and inference pipelines of the proposed method in Section \ref{section Training and Inference}. The overall framework of our approach is illustrated in Figure \ref{fig:FAIR}.

\subsection{Problem Formulation}
\label{section Problem Formulation}
Given an item corpus \( I \) and a user’s historical interaction sequence \( U = [u_1, u_2, \dots, u_{t-1}] \) where \( u \in I \), the target of a sequential recommendation system is to predict  the next most likely item \( u_t \in I \) that the user may interact with.

In our approach, to achieve better representational capacity, we discretize each item into multiple codes \( C = [c^1, c^2, \dots, c^L]  \), where $c^i \in \mathcal{C}^i \text{ for } i \in [1,2,...,L]$ and \( L \) ($L \gg 4$) denotes the length of the code sequence, in contrast to the standard practice \cite{rajput2023recommender} in generative recommendation that uses only four codes per item. Given that $L$ is relatively large, and in consideration of computational efficiency and resource constraints, we adopt a non-autoregressive (NAR) modeling strategy \cite{ren2020study} in conjunction with the \textit{multi-token prediction} objective \cite{gloeckle2024better}, which predicts all codes of the next item in a single forward pass. The corresponding loss function is formulated as follows:
\begin{equation}\label{eq:mtp loss}
\begin{aligned}
\mathcal{L}_{\text{MTP}} 
&= -\log p(u_t \mid u_1, u_2, \dots, u_{t-1}) \\
&= - \sum_{k=1}^{L} \log p(c_t^k \mid u_1, u_2, \dots, u_{t-1})
\end{aligned}
\end{equation}

\subsection{Focused Attention Mechanism}
\label{section Focused Attention}
Given an input sequence $X \in \mathbb{R}^{n \times d}$, where $n$ denotes the sequence length and $d$ denotes the embedding dimension, the focused attention mechanism maps the input into a set of query, key, and value vectors. Specifically, the input is linearly projected through distinct learnable matrices:
\begin{align}
Q_i &= X W_{Q_i}, \quad K_i = X W_{K_i}, \quad V = X W_V, \quad i \in \{1,2\} \label{eq:proj}
\end{align}
where $W_{Q_i}, W_{K_i} \in \mathbb{R}^{d \times d_k}$, $W_V \in \mathbb{R}^{d \times d_v}$, and $d_k$ is the dimensionality of queries and keys.  

For each branch $i$, the scaled dot-product attention scores are computed as:
\begin{align}
A_i &= \text{Softmax}\!\left(\frac{Q_i K_i^\top}{\sqrt{d_k}}\right), \quad i \in \{1,2\} \label{eq:attn_scores}
\end{align}

Unlike the classical attention mechanism \cite{vaswani2017attention} which uses a single attention matrix, focused attention explicitly models the difference between the two attention distributions to \textit{eliminate attention noise}. The focused attention matrix is defined as:
\begin{align}
A &= \text{Norm}(\lambda_1 A_1 - \lambda_2 A_2) \label{eq:diff_matrix}
\end{align}
where $\lambda_1$ and $\lambda_2$ are hyperparameters to adjust the relative importance of the two attention distributions, and Norm indicates the normalization operation.

The final output is then obtained by applying this focused attention matrix to the value vectors:
\begin{align}
\text{F-Attn}(X) &= A V \label{eq:diffattn}
\end{align}

\begin{figure*}[htbp]
  \centering
  \includegraphics[width=\linewidth]{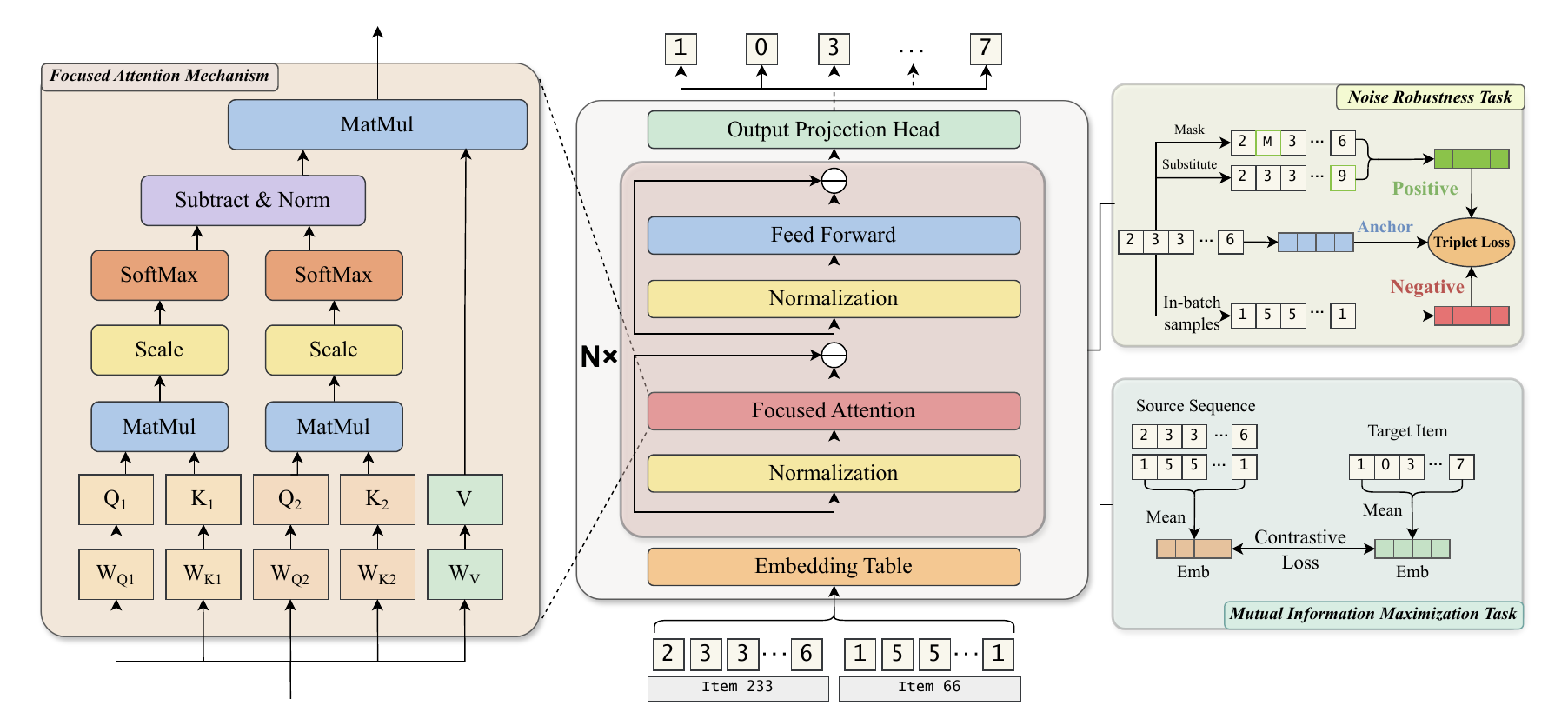}
  \caption{An overview of FAIR. FAIR consists of three major components: Focused Attention Mechanism (FAM), Noise-Robustness Task (NRT), and Mutual Information Maximization Task (MIM).}
  \label{fig:FAIR}
\end{figure*}

This subtraction operation suppresses noisy or redundant correlations captured by one branch while emphasizing more informative dependencies highlighted by the other, thereby improving the robustness and discrimination ability of the attention mechanism as shown in Figure \ref{fig:FAIR}.  

To further enrich the representational capacity, the mechanism can be extended to a multi-head setting, analogous to multi-head attention. Specifically, the input $X$ is projected into $h$ different subspaces, and focused attention is applied independently in each head:
\begin{align}
\text{head}_j &= \text{F-Attn}_j\!\left(X \right), \quad j = 1, \dots, h \label{eq:multihead_head}
\end{align}

The outputs from all heads are concatenated and linearly transformed to form the final representation:
\begin{align}
\text{MultiHead F-Attn}(X) &= \text{Concat}(\text{head}_1, \dots, \text{head}_h) W_O \label{eq:multihead_output}
\end{align}
where $W_O \in \mathbb{R}^{hd_v \times d}$.

This extension allows the model to simultaneously capture diverse signals from multiple representation subspaces, further enhancing its ability to distinguish relevant information from noise.

\subsection{Noise-Robustness Task}
\label{section Noise-Robustness Task}
To enhance the robustness of the learned representations, we introduce a noise-robustness objective. The key idea is to simulate noisy inputs by randomly perturbing parts of the original codes sequence, and then enforce strong consistency constraints such that the hidden representations of the noisy and clean inputs remain aligned.

Formally, let the original codes sequence be
\[
\mathbf{c} = [c_1, c_2, \dots, c_L],
\]
where $c_i \in \mathcal{C}^i$ is drawn from the corresponding codebook $\mathcal{C}^i$. We construct a noisy version $\tilde{\mathbf{c}}$ by perturbing $\mathbf{c}$ according to the following stochastic rule:
\begin{align}
\tilde{c}_i = 
\begin{cases}
\texttt{[PAD]}, & \text{with probability } p_{\text{mask}}, \\
c, & \text{with probability } p_{\text{sub}}, \, c \sim \mathcal{U}(\mathcal{C}^i \setminus\{c_i\}), \\
c_i, & \text{otherwise},
\end{cases}
\end{align}
where $p_{\text{mask}}$ and $p_{\text{sub}}$ denote the probabilities of masking and substitution, respectively, and $\mathcal{U}(\mathcal{C}^i\setminus\{c_i\})$ represents a uniform draw from the same codebook excluding $c_i$. In practice, we set $p_{\text{mask}}=0.1$ and $p_{\text{sub}}=0.1$.

Let $h$ and $\tilde{h}$ denote the hidden representations produced by the model for the original and noisy codes sequence, respectively. To encourage robustness, we require $\mathbf{h}$ and $\tilde{\mathbf{h}}$ to remain close, while separating them from negative samples $\mathbf{h}^-$ drawn from other items within the same batch. This is achieved via a triplet loss \cite{schroff2015facenet}:
\begin{align}
\mathcal{L}_{\text{NR}}
= \frac{1}{N} \sum
\max \Big(0, \, d(\mathbf{h}, \tilde{\mathbf{h}}) - d(\mathbf{h}, \mathbf{h}^-) + m \Big),
\end{align}
where $d(\cdot,\cdot)$ is a distance metric (e.g., cosine distance), $m$ is a margin hyperparameter and $N$ is the number of negative samples.

By minimizing $\mathcal{L}_{\text{NR}}$, the model learns to be invariant to random perturbations in the input codes, thereby improving robustness against noisy or irrelevant tokens.

\subsection{Mutual Information Maximization Task}
\label{section Mutual Information Maximization Task}
To suppress attention noise and make the model focus on informative contextual signals, we introduce a mutual information maximization (MIM) objective. As illustrated in Figure \ref{fig:FAIR}, we obtain compact representations by mean-pooling embeddings of the source sequence and of the target item, then apply a contrastive loss that brings the positive target--sequence pair closer while pushing apart negatives drawn from other sequences. The goal is to maximize the dependency between the two representations so that the model captures information useful for prediction.

Directly maximizing the mutual information 
\begin{align}
I(\mathbf{h}_t;\mathbf{x}_t) = \mathbb{E}_{\mathbf{h}_t, \mathbf{x}_t} \Big[ \log \frac{p(\mathbf{h}_t, \mathbf{x}_t)}{p(\mathbf{h}_t)p(\mathbf{x}_t)} \Big]
\end{align}
is challenging due to the intractability of the joint and marginal distributions. To address this, we adopt a contrastive surrogate, namely the InfoNCE objective \cite{chen2020simple}. Given a target embedding $\mathbf{h}_t$, a context embedding $\mathbf{x}_t$, and a set of negatives $\{\mathbf{x}_t^-\}$ from the same batch, the loss is defined as
\begin{align}
\mathcal{L}_{\text{MIM}}
= - \log \frac{\exp\big(\mathrm{sim}(\mathbf{h}_t, \mathbf{x}_t)/\tau\big)}
{\exp\big(\mathrm{sim}(\mathbf{h}_t, \mathbf{x}_t)/\tau\big) + \sum\exp\big(\mathrm{sim}(\mathbf{h}_t, \mathbf{x}_t^-)/\tau\big)}
\end{align}
where $\mathrm{sim}(\cdot, \cdot)$ denotes cosine similarity. 

The connection to mutual information can be seen by viewing InfoNCE as a $K$-way classification problem. Let $\mathcal{X} = \{\mathbf{x}_1, \dots, \mathbf{x}_K\}$ denote the candidate context embeddings. The true posterior that $\mathbf{x}_i$ is the positive sample is
\begin{align}
p(i \mid \mathbf{h}_t, \mathcal{X}) = \frac{p(\mathbf{x}_i \mid \mathbf{h}_t)/p(\mathbf{x}_i)}{\sum_{j=1}^{K} p(\mathbf{x}_j \mid \mathbf{h}_t)/p(\mathbf{x}_j)}
\end{align}
while the model predicts
\begin{align}
p_\theta(i \mid \mathbf{h}_t, \mathcal{X}) = \frac{\exp(\mathrm{sim}(\mathbf{h}_t, \mathbf{x}_i)/\tau)}{\sum_{j=1}^{K} \exp(\mathrm{sim}(\mathbf{h}_t, \mathbf{x}_j)/\tau)}
\end{align}

Minimizing the InfoNCE loss is equivalent to minimizing the cross-entropy between $p$ and $p_\theta$:
\begin{equation}
\begin{aligned}
\mathcal{L}_{\text{InfoNCE}} 
&= - \sum_{i=1}^{K} p(i \mid \mathbf{h}_t, \mathcal{X}) \log p_\theta(i \mid \mathbf{h}_t, \mathcal{X}) \\
&= H(p) + \mathrm{KL}(p \| p_\theta),
\end{aligned}
\end{equation}
where $H(p)$ is the entropy of the true posterior and $\mathrm{KL}(p \| p_\theta) \ge 0$. Since the KL term is non-negative, minimizing $\mathcal{L}_{\text{InfoNCE}}$ effectively maximizes the log-likelihood of the positive pair.

Furthermore, using \textit{Jensen's inequality} \cite{jensen1906fonctions}, we can upper-bound the negative-sample contribution and relate the loss to mutual information:
\begin{align}
\log \sum_{j=1}^{K} \frac{p(\mathbf{x}_j \mid \mathbf{h}_t)}{p(\mathbf{x}_j)}
\le \sum_{j=1}^{K} \frac{1}{K} \log \frac{p(\mathbf{x}_j \mid \mathbf{h}_t)}{p(\mathbf{x}_j)} + \varepsilon_K
\end{align}
where $\varepsilon_K$ is a small term due to finite negatives. This leads to the mutual information lower bound:
\begin{align}
I(\mathbf{h}_t; \mathbf{x}_t) \ge \log K - \mathcal{L}_{\text{MIM}} - \varepsilon_K
\end{align}

In summary, the MIM task encourages the target representation to be predictive of its true context while being distinct from unrelated ones, reducing attention to irrelevant tokens and reinforcing informative dependencies.

\subsection{Training and Inference}
\label{section Training and Inference}
\subsubsection{Training} 
We combine the multi-token prediction, noise-robustness and mutual information maximization losses to train our model, given by:
\begin{align}
    \mathcal{L} = \mathcal{L}_{\text{MTP}} + \alpha \mathcal{L}_{\text{NR}} + \beta \mathcal{L}_{\text{MIM}}
\end{align}
where $\alpha$ and $\beta$ are loss coefficients.

\subsubsection{Inference} 
During inference phase, beam search is applied to generate all codes in parallel. Once finished, the codes are mapped to their corresponding items through the lookup table, creating a ranked recommendation list based on confidence scores to produce the final top-$k$ results.

\section{Experiments}
We analyze the proposed FAIR method on four datasets and
demonstrate its effectiveness by answering the following research
questions:

\begin{itemize}[left=0pt]
\item \textbf{RQ1}: How does FAIR perform compared with existing best-performing sequential recommendation methods among different datasets?
\item \textbf{RQ2}: Do the components (e.g., focused attention, noise-rob\\ustness task and mutual information maximization task) of FAIR contribute positively to its performance?
\item \textbf{RQ3}: How do hyper-parameter settings affect the performance of FAIR?
\end{itemize}

\begin{table}[t]
\renewcommand{\arraystretch}{1.2}
    \centering
    \caption{Statistics of the Datasets.}
    \label{tab:Statistics of the Datasets}
    \resizebox{\linewidth}{!}{
    \begin{tabular}{ccccc}
        \toprule
        Dataset & \#Users & \#Items & \#Interactions & \#Density \\
        \midrule
        Beauty & 22,363 & 12,101 & 198,360 & 0.00073 \\
        Sports and Outdoors & 35,598 & 18,357 & 296,175 & 0.00045 \\
        Toys and Games & 19,412 & 11,924 & 167,526 & 0.00073 \\
        CDs and Vinyl & 75,258 & 64,443 & 1,022,334 & 0.00021 \\
        \bottomrule
    \end{tabular}
    }
\end{table}

\subsection{Experimental Setting}
\subsubsection{Dataset.}
We conduct experiments on four public benchmarks commonly used in the sequential recommendation task. For all datasets, we group the interaction records by users and sort them by the interaction timestamps ascendingly. Following \cite{rendle2010factorizing, zhang2019feature}, we only keep the 5-core dataset, which filters unpopular items and inactive users with fewer than five interaction records. Statistics of these datasets are shown in Table \ref{tab:Statistics of the Datasets}.

\begin{itemize}[left=0pt]
\item \textbf{Amazon}: Amazon Product Reviews dataset \cite{mcauley2015image}, containing user reviews and item metadata from May 1996 to July 2014. In particular, we use four categories of the Amazon Product Reviews dataset for the sequential recommendation task: "Beauty", "Sports and Outdoors", "Toys and Games" and “CDs and Vinyl". In addition, we include “CDs and Vinyl" category, which has approximately five times more interactions than “Beauty” category, to evaluate performance on larger datasets.
\end{itemize}

\subsubsection{Evaluation Metrics.}
We employ two broadly used criteria for the matching phase, \textit{i.e.}, Recall and Normalized Discounted Cumulative Gain (NDCG). We report metrics computed on the top 5/10 recommended candidates. Following the standard evaluation protocol \cite{kang2018self}, we use the leave-one-out strategy for evaluation. For each item sequence, the last item is used for testing, the item before the last is used for validation, and the rest is used for training.

\subsubsection{Implementation Details.}
We implement our method using HuggingFace transformers \cite{wolf2020transformers} based decoder-only architecture. Specifically, we use a 2-layer Transformer decoder with an embedding dimension of $d=512$, a feed-forward layer dimension of 1024, and 4 attention heads. We use the SiLU activation function \cite{elfwing2018sigmoid} and RMS Normalization \cite{zhang2019root} for all the layers. For item codes, following previous practice \cite{rajput2023recommender}, we use sentence-t5-base \cite{ni2021sentence} semantic encoder to extract item semantic embeddings and optimized product quantization (OPQ) \cite{ge2013optimized} to produce 32 codes for each item. For focused attention module, we set both attention coefficients $\lambda_1$ and $\lambda_2$ to 1. For noise-robustness task, we set the margin hyperparameter $m$ to 0.2. For mutual information maximization task, we set the temperature $\tau$ to 0.03. To train our model, we adopt Adam optimizer with the learning rate 0.003, and employ the warmup strategy for stable training. Besides, the loss coefficients $\alpha$ and $\beta$ are set to $1$ and $0.01$ respectively. 
For additional details and analyses, please refer to Appendix \ref{Appendix:details}.

\begin{table*}[htbp]
    \renewcommand{\arraystretch}{1.5}
    \centering
    \caption{Performance comparison of different methods. The best performance is highlighted in bold while the second best performance is underlined. All the results of FAIR are statistically significant with $p < 0.05$ compared to the best baseline models by paired t-test.}
    \label{tab:overall performance}
    \resizebox{\linewidth}{!}{
    \begin{tabular}{clccccccccccccc}
        \toprule
        \multirow{2}{*}{\textbf{Dataset}} & \multirow{2}{*}{\textbf{Metric}} & \multicolumn{7}{c}{\textbf{Traditional}} & \multicolumn{5}{c}{\textbf{Generative}} & \multirow{2}{*}{\textbf{FAIR (Ours)}} \\
        \cmidrule(lr){3-9} \cmidrule{10-14}
        & & \textbf{GRU4REC} & \textbf{Caser} & \textbf{SASRec} & \textbf{BERT4Rec} & \textbf{HGN} & \textbf{FDSA} & \textbf{S$^3$-Rec} & \textbf{RecJPQ} & \textbf{VQ-Rec} & \textbf{TIGER} & \textbf{HSTU} & \textbf{RPG} \\
        \midrule
        \multirow{4}{*}{\textbf{Beauty}} 
        & Recall@5 & 0.0164 & 0.0205 & 0.0387 & 0.0203 & 0.0325 & 0.0267 & 0.0387 & 0.0311 & 0.0457 & 0.0454 & 0.0469 & \underline{0.0504} & \textbf{0.0563} \\
        & Recall@10 & 0.0283 & 0.0347 & 0.0605 & 0.0347 & 0.0512 & 0.0407 & 0.0647 & 0.0482 & 0.0664 & 0.0648 & 0.0704 & \underline{0.0730} & \textbf{0.0783}  \\
        & NDCG@5   & 0.0099 & 0.0131 & 0.0249 & 0.0124 & 0.0206 & 0.0206 & 0.0244 & 0.0167 & 0.0317 & 0.0321 & 0.0314 & \underline{0.0349} & \textbf{0.0395} \\
        & NDCG@10  & 0.0137 & 0.0176 & 0.0318 & 0.0170 & 0.0266 & 0.0266 & 0.0327 & 0.0222 & 0.0383 & 0.0384 & 0.0389 & \underline{0.0422} & \textbf{0.0465}  \\
        \midrule
        \multirow{4}{*}{\textbf{Sports}} 
        & Recall@5 & 0.0129 & 0.0116 & 0.0233 & 0.0115 & 0.0189 & 0.0182 & 0.0251 & 0.0141 & 0.0208 & 0.0264 &  0.0258 & \underline{0.0294} & \textbf{0.0317}  \\
        & Recall@10 & 0.0204 & 0.0194 & 0.0350 & 0.0191 & 0.0313 & 0.0288 & 0.0385 & 0.0220 & 0.0300 & 0.0400 & 0.0414 & \underline{0.0419} &  \textbf{0.0455}   \\
        & NDCG@5 & 0.0086 & 0.0072 & 0.0154 & 0.0075 & 0.0120 & 0.0122 & 0.0161 & 0.0076 & 0.0144 & 0.0181 &  0.0165 & \underline{0.0201} & \textbf{0.0224} \\
        & NDCG@10 & 0.0110 & 0.0097 & 0.0192 & 0.0099 & 0.0159 & 0.0156 & 0.0204 & 0.0102 & 0.0173 & 0.0225 & 0.0215 & \underline{0.0241} & \textbf{0.0268} \\
        \midrule
        \multirow{4}{*}{\textbf{Toys}} 
        & Recall@5 & 0.0097 & 0.0166 & 0.0463 & 0.0116 & 0.0321 & 0.0228 & 0.0443 & 0.0331 & 0.0497 & 0.0521 & 0.0433 & \underline{0.0531} & \textbf{0.0601}  \\
        & Recall@10 & 0.0176 & 0.0270 & 0.0675 & 0.0203 & 0.0497 & 0.0381 & 0.0700 & 0.0484 & 0.0737 & 0.0712 & 0.0669 & \underline{0.0759} & \textbf{0.0836}  \\
        & NDCG@5 & 0.0059 & 0.0107 & 0.0306 & 0.0071 & 0.0221 & 0.0140 & 0.0294 & 0.0182 & 0.0346 & 0.0371 & 0.0281 &\underline{0.0373} & \textbf{0.0414} \\
        & NDCG@10 & 0.0084 & 0.0141 & 0.0374 & 0.0099 & 0.0277 & 0.0189 & 0.0376 & 0.0231 & 0.0423 & 0.0432 & 0.0357 & \underline{0.0446} & \textbf{0.0490} \\
        \midrule
        \multirow{4}{*}{\textbf{CDs}}
        & Recall@5 & 0.0195 & 0.0116 & 0.0351 & 0.0326 & 0.0259 & 0.0226 & 0.0213 & 0.0075 & 0.0352 & \underline{0.0492} & 0.0417 & 0.0482 & \textbf{0.0520}  \\
        & Recall@10 & 0.0353 & 0.0205 & 0.0619 & 0.0547 & 0.0467 & 0.0378 & 0.0375 & 0.0138 & 0.0520 & \textbf{0.0748} & 0.0638 & 0.0720 & \underline{0.0739}  \\
        & NDCG@5 & 0.0120 & 0.0073 & 0.0177 & 0.0201 & 0.0153 & 0.0137 & 0.0130 & 0.0046 & 0.0238 & \underline{0.0329} & 0.0275 & 0.0325 & \textbf{0.0373} \\
        & NDCG@10 & 0.0171 & 0.0101 & 0.0263 & 0.0271 & 0.0220 & 0.0186 & 0.0182 & 0.0066 & 0.0292 & \underline{0.0411} & 0.0346 & 0.0402 & \textbf{0.0431} \\
        \bottomrule
    \end{tabular}}
\end{table*}

\subsection{Performance Comparison (RQ1)}
\subsubsection{Baselines.}
We compare FAIR with several representative related baselines, encompassing traditional sequential modeling approaches as well as the latest emergent generative techniques.

\noindent (1) For \textit{traditional} methods, we have:
\begin{itemize}[left=0pt]
    \item \textbf{GRU4REC} \cite{hidasi2015session} is an RNN-based model that utilizes Gated Recurrent Units (GRUs) to model user click sequences.
    \item \textbf{Caser} \cite{tang2018personalized} is a CNN-based method capturing high-order Markov Chains by modeling user behaviors through both horizontal and vertical convolutional operations.
    \item \textbf{SASRec} \cite{kang2018self} is a self-attention-based sequential recommendation model that utilizes a unidirectional Transformer encoder with a multi-head attention mechanism to effectively model user behavior and predict the next item in a sequence.
    \item \textbf{BERT4Rec} \cite{sun2019bert4rec} adopts a Transformer model with the bidirectional self-attention mechanism and uses a Cloze objective loss for the modeling of item sequences.
    \item \textbf{HGN} \cite{ma2019hierarchical} adopts hierarchical gating networks to capture long-term and short-term user interests.
    \item \textbf{FDSA} \cite{zhang2019feature} leverages self-attention networks to separately model item-level and feature-level sequences, emphasizing the transformation patterns between item features and utilizing a feature-level self-attention block to capture feature transition dynamics.
    \item \textbf{S$^3$-Rec} \cite{zhou2020s3} employs mutual information maximization to pre-train a bi-directional Transformer for sequential recommendation, enhancing the model's ability to learn correlations between items and their attributes through self-supervised tasks.
\end{itemize}

\noindent (2) For \textit{generative} methods, we have:
\begin{itemize}[left=0pt]
    \item \textbf{RecJPQ} \cite{petrov2024recjpq} adopts joint product quantization to compress item embeddings into a small set of shared sub-embeddings.
    \item \textbf{VQRec} \cite{hou2023learning} maps item text into discrete codes via vector quantization, then derives item representations from these codes, enhancing recommendation performance through contrastive pre-training with hard negatives and a differentiable permutation-based fine-tuning strategy.
    \item \textbf{TIGER} \cite{rajput2023recommender} utilizes a pre-trained T5 encoder to learn semantic identifiers for items, autoregressively decodes target candidates using these identifiers, and incorporates RQ-VAE quantization to build generalized item identifiers without relying on sequential order.
    \item \textbf{HSTU} \cite{zhai2024actions} reformulates recommendation as a generative modeling task by discretizing raw item features into tokens and processing them with a high-efficiency architecture tailored for industrial recommendation data.
     \item \textbf{RPG} \cite{hou2025generating} employs a multi-token prediction training objective and utilizes unordered long semantic IDs that provide expressive item representations, all while preserving efficient inference through parallel prediction.

\end{itemize}

\begin{table*}[t]
\renewcommand{\arraystretch}{1.5}
\centering
\caption{Ablation studies by selectively discarding the focused Attention Mechanism (FAM), Noise-Robustness Task (NRT) and Mutual Information Maximization Task (MIM).}
\begin{tabular}{ccccccccccc}
\toprule
\multicolumn{3}{c}{Variants} & \multicolumn{4}{c}{Beauty} & \multicolumn{4}{c}{Toys} \\ \cmidrule(lr){1-3} \cmidrule(lr){4-7} \cmidrule(lr){8-11}
FAM & NRT & MIM & Recall@5 & NDCG@5 & Recall@10 & NDCG@10 & Recall@5 & NDCG@5 & Recall@10 & NDCG@10\\ 
\midrule
 &  &  & 0.0490  & 0.0353  & 0.0723  & 0.0429  & 0.0503  & 0.0353  & 0.0730  & 0.0427  \\
\rowcolor{linecolor1} \checkmark &  &  & 0.0514 & 0.0361 & 0.0763 & 0.0442 & 0.0535  & 0.0376  & 0.0770  & 0.0452  \\
\rowcolor{linecolor2} \checkmark & \checkmark &  & 0.0546 & 0.0386 & 0.0777 & 0.0457 & 0.0575  & 0.0395  & 0.0819  & 0.0473  \\
\rowcolor{linecolor} \checkmark & \checkmark & \checkmark & 0.0563 & 0.0395 & 0.0783 & 0.0465 & 0.0601 & 0.0414 & 0.0836 & 0.0490 \\ \bottomrule
\end{tabular}  \label{tab:ablation study}
\end{table*}

\subsubsection{Overall Performance.} Table \ref{tab:overall performance} reports the overall performance of four datasets. The results for most of baselines are taken from the publicly accessible results \cite{zhou2020s3, hou2025generating}\footnote{https://github.com/RUCAIBox/CIKM2020-S3Rec}. 
For RPG, we modify the semantic encoder to sentence-t5-base\cite{ni2021sentence} for fair comparison, while keeping other configurations consistent with its official implementation\footnote{RPG employs OpenAI’s text-embedding-3-large API, which provides substantially stronger representation capabilities compared with the sentence-t5-base encoder used in other baselines. This difference may introduce an advantage unrelated to model design. The official implementation of RPG is publicly available at: https://github.com/facebookresearch/RPG\_KDD2025}.

From the results, we have the following observations:
\begin{itemize}[left=0pt]
    \item \textbf{Generative methods typically achieve superior performance compared to traditional approaches.} Conventional methods often rely on arbitrary numerical identifiers that lack semantic meaning, whereas generative methods pre-encode semantic side information of items into discrete codes, resulting in more informative item representations. Furthermore, traditional approaches identify top-k candidate items through inner-product similarity, a process whose computational cost increases significantly as the item set grows. In contrast, generative methods can directly generate the top-k candidates via beam search, with a computational complexity that is largely decoupled from the size of the item set. This property highlights their potential to replace the cascaded architectures commonly used in industrial recommendation systems.

    \item \textbf{FAIR consistently outperforms nearly all baseline models across multiple datasets, achieving top ranking on 15 out of 16 metrics.} Notably, on the Toys benchmark, FAIR improves Recall@5 by $13.18\%$ and NDCG@5 by $10.99\%$ compared to the strongest baseline, RPG.
    These significant improvements can be attributed to our novel generative framework, which aims to mitigate attention noise, a common yet often overlooked issue in generative recommendation systems. The framework endows the model's ability to focus on truly relevant user-item interactions while filtering out spurious correlations.

    \item \textbf{FAIR also demonstrates superior performance over previous generative approaches across most datasets. } 
    This advantage stems from three key innovations: its enhanced capability to discriminate decision-relevant contexts from noise, significantly strengthened robustness against various types of input perturbations, and the incorporation of a focused attention mechanism that adaptively focuses on relevant contexts while suppressing noise. 
    These significant improvements demonstrate the effectiveness of our approach and underscore the importance of addressing attention noise issue in generative recommendation  systems.

\end{itemize}

\begin{figure*}[htbp]
    \centering
    \includegraphics[width=0.9\linewidth]{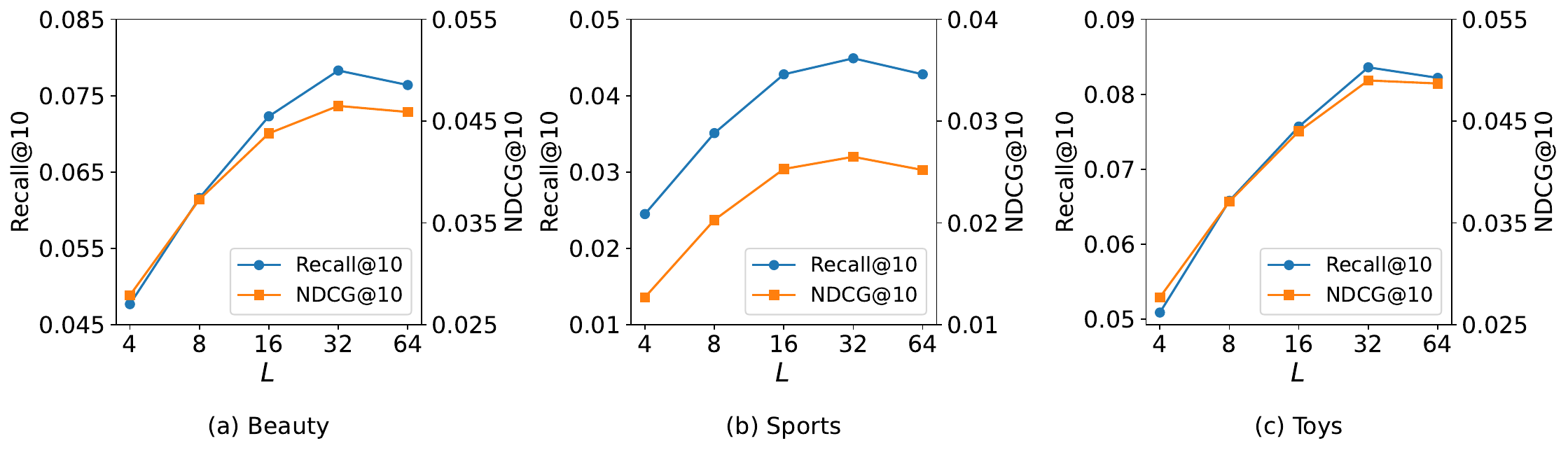}
    \caption{Analysis of the performance impact of the length of code sequences $L$.}
    \label{fig:code length}
\end{figure*}

\begin{figure*}[htbp]
    \centering
    \includegraphics[width=0.9\linewidth]{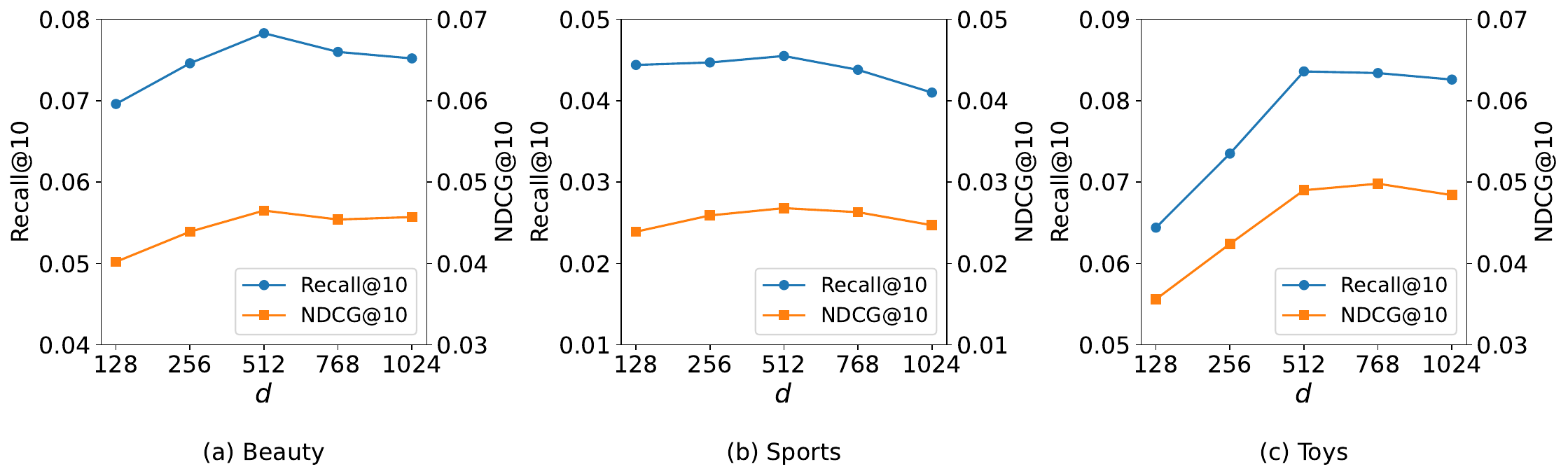}
    \caption{Analysis of the performance impact of the embedding dimension of model $d$.}
    \label{fig:emb dim}
\end{figure*}

\subsection{Ablation Study (RQ2)}
We evaluated the performance impact of FAIR’s components via an ablation study. Specifically, we gradually discard the focused attention mechanism (FAM), noise-robustness task (NRT) and mutual information maximization task (MIM) from FAIR to obtain ablation architectures. The results are reported in Table \ref{tab:ablation study}, we can observe that:




\begin{itemize}[left=0pt]
    \item Removing any FAM, NRT or MIN leads to performance degradation, with removal of all components yielding the worst results across different datasets. These findings highlight the inherent superiority and robustness of our framework in mitigating attention noise issue for generative recommendation.

    \item Removing FAM causes notable drops, confirming the necessity of suppressing irrelevant contexts. FAM effectively highlights informative signals while attenuating interference, thereby enhancing recommendation performance.

    \item Removing either NRT or MIM results in substantial performance decline. NRT improves robustness by exposing the model to random perturbations, while MIM is indispensable for capturing most informative contexts, together ensuring stable and predictive representations for next-item prediction.
\end{itemize}

\subsection{Further Analysis (RQ3)}

\subsubsection{The Length of Code Sequences L}
We conducted a systematic investigation into the impact of varying the code length $L$ on model performance. A larger value of $L$ corresponds to a higher capacity representation, allowing the model to encode more fine grained semantic information into the discrete codes. Experiments were carried out across three datasets, including Beauty, Sports, and Toys, with results summarized in Figure \ref{fig:code length}.

Our observations indicate that as $L$ increases from 4 to 32, the model performance shows continuous improvement across all datasets. This trend underscores the advantage of longer codes in enhancing semantic expressiveness, which we attribute to their ability to capture richer and more nuanced item characteristics. However, when $L$ is further increased to 64, performance on all datasets begins to plateau or even experience a slight decline. This suggests that while longer codes offer improved representational capacity, there exists a threshold beyond which additional code length may introduce redundancy, complicate the learning process, and increase optimization difficulty.

In summary, these findings emphasize the importance of selecting an appropriate code length. Although longer codes help enrich the semantic representation space, excessively large values of $L$ can impair both training efficiency and generalization performance. In our experimental setting, $L = 32$ emerges as the optimal configuration, achieving an effective balance between representational power and practical feasibility.

\subsubsection{The Embedding Dimension of Model $d$}
We conducted a systematic study to evaluate the impact of embedding dimension $d$ on model performance. Generally, a larger $d$ enhances the model's capacity to capture intricate semantic information, but it also increases the risk of overfitting and computational overhead. Experiments were performed on three datasets: Beauty, Sports, and Toys, with the results illustrated in Figure \ref{fig:emb dim}.

We observe that as $d$ increases from 128 to 512, performance improves consistently across all datasets, indicating that higher-dimensional embeddings contribute to greater representational power and improved semantic expressiveness. However, when $d$ is further increased to 768 and 1024, performance begins to plateau or even degrade, with the most noticeable decline on the Sports dataset. This reversal suggests that overly large embedding dimensions may introduce redundant information, complicate the training process, and ultimately impair generalization capability.

In summary, these results underscore the importance of selecting an appropriate embedding size to balance representational capacity and practical efficiency. Based on our empirical findings, a dimension of $d=512$ achieves the optimal trade-off between expressiveness and robustness in our experiments.

\subsection{Case Study}
\begin{table}[htbp]
\renewcommand{\arraystretch}{1.5}
\centering
\caption{The impact of individual component of FAIR on attention allocation. The top row shows “History” for input items and “Target” for the target item. The row under each history item shows whether the item is relevant context (\textcolor{green}{\checkmark}) or noise (\textcolor{red}{\ding{55}}). Numbers indicate attention weights; arrows show change relative to the previous step. The emoji reflects the rationality of the attention distribution: a sad face indicates unreasonable attention dominated by noise,  a puzzled face suggests partially-correct attention,  and a smiling face represents fully aligned and meaningful attention.}
\begin{tabular}{ccc c c c c c}
\toprule
\multicolumn{3}{c}{Variants} & \multicolumn{4}{c}{History} & Target \\
\cmidrule(lr){1-3} \cmidrule(lr){4-7} \cmidrule(lr){8-8}
\makecell{FAM} & 
\makecell{NRT} & 
\makecell{MIM} &
\makecell{\includegraphics[width=0.5cm]{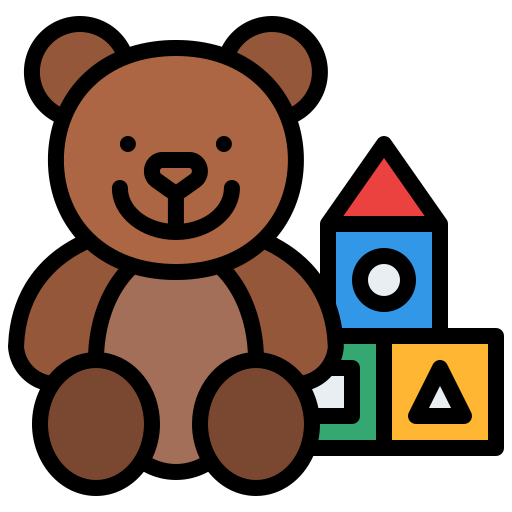} \\ \textcolor{red}{\ding{55}}} & 
\makecell{\includegraphics[width=0.5cm]{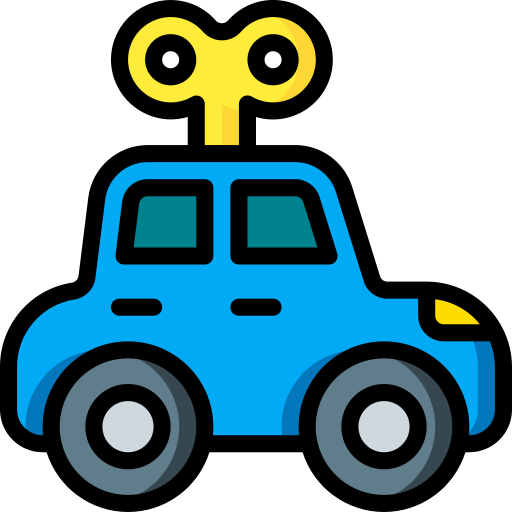} \\  \textcolor{green}{\checkmark}} & 
\makecell{\includegraphics[width=0.5cm]{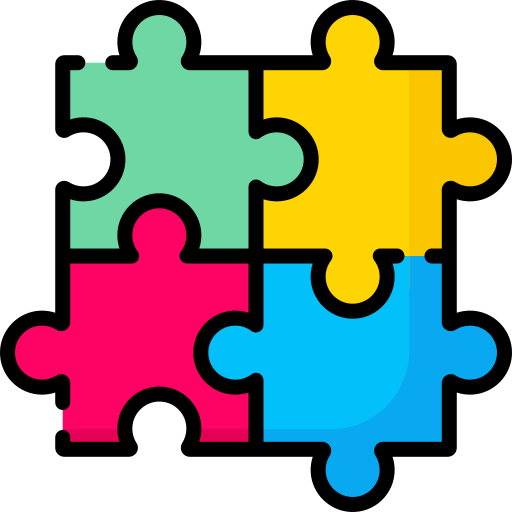} \\ \textcolor{red}{\ding{55}}} & 
\makecell{\includegraphics[width=0.5cm]{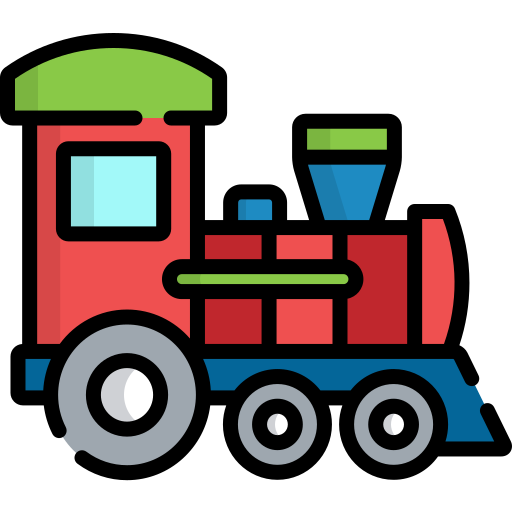} \\ \textcolor{green}{\checkmark}} & 
\makecell{\includegraphics[width=0.5cm]{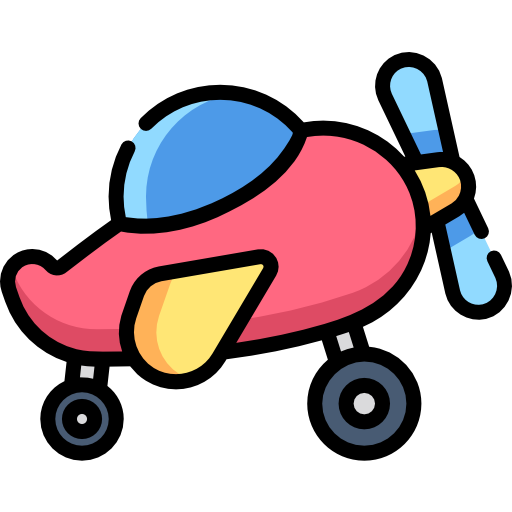}} \\

\midrule
 &  &  & 0.27 & 0.18 & 0.33 & 0.22 & \makecell{
  \centering
  \includegraphics[width=0.6cm]{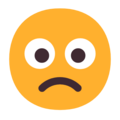}
} \\ 
\checkmark &  &  & 0.30 \textcolor{green}{↑} & 0.24 \textcolor{green}{↑} & 0.20 \textcolor{red}{↓} & 0.26 \textcolor{green}{↑} & \makecell{
  \centering
  \includegraphics[width=0.6cm]{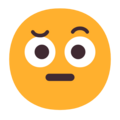}
} \\ 
\checkmark & \checkmark &  & 0.33 \textcolor{green}{↑} & 0.16 \textcolor{red}{↓} & 0.14 \textcolor{red}{↓} & 0.37 \textcolor{green}{↑} & \makecell{
  \centering
  \includegraphics[width=0.6cm]{./figures/puzzle.png}
} \\ 
\checkmark & \checkmark & \checkmark & 0.12 \textcolor{red}{↓} & 0.38 \textcolor{green}{↑} & 0.08 \textcolor{red}{↓} & 0.42 \textcolor{green}{↑} & \makecell{
  \centering
  \includegraphics[width=0.6cm]{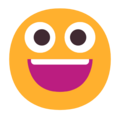}
} \\ 
\bottomrule
\end{tabular}  
\label{tab:case study}
\end{table}



To investigate the impact of individual components in our FAIR framework on attention allocation, we conduct a series of ablation studies by progressively removing the Focused Attention Mechanism (FAM), Noise-Robustness Task (NRT), and Mutual Information Maximization (MIM) module through a more fine-grained analysis of the user case study depicted in Figure \ref{fig:case study}. We then analyze the resulting attention patterns across these ablated configurations, as summarized in Table \ref{tab:case study}. Our analysis indicates that attention refinement does not follow a monotonic trajectory.

The inclusion of FAM leads to heightened attention on relevant objects—for instance, the attention score for “Toy Train” increases from 0.22 to 0.26. However, FAM can also inadvertently amplify attention to noise, as observed with “Toy Bear,” which rises from 0.27 to 0.30. Introducing the NRT steers attention toward contextual features that remain stable under perturbation. This often suppresses attention to superficially relevant but unstable items, such as “Toy Car,” whose score declines from 0.24 to 0.16, while more persistent noise elements like “Toy Bear” may persist or even slightly increase (to 0.33).

It is only with the integration of MIM that attention becomes fully aligned with the most predictive contexts. Under the complete framework, relevant items such as “Toy Car” and “Toy Train” achieve attention scores of 0.38 and 0.42, respectively, while noisy distractors like “Toy Bear” and “Puzzle” are effectively suppressed to 0.12 and 0.08. These findings demonstrate that FAM enhances feature discriminability, NRT enforces robustness against perturbations, and MIM consolidates attention toward the most informative contexts for predicting the target “Toy Plane”.

\section{Conclusion}
In this work, we propose a novel framework named FAIR for the first time to endow transformer-based generative models with the ability to enhance attention to relevant context while suppressing noise. FAIR comprises three key components: (1) a focused attention mechanism, which learns two distinct sets of query and key matrices to compute the attention through subtraction to highlight relevant context. (2) a noise-robustness task that improves he model’s robustness to irrelevant noise. (3) a mutual information maximization task that  encourages the extraction of contexts that are most informative for enhancing next item recommendation. Extensive comparisons with state-of-the-art methods and detailed analyses demonstrate the effectiveness and robustness of FAIR.







\bibliographystyle{ACM-Reference-Format}
\bibliography{reference}

\clearpage

\appendix
\begin{center}
    {\Large \textbf{Appendix}} 
\end{center}

\section{Additional Implementation Details}
\label{Appendix:details}
\begin{table}[htbp]
\centering
\caption{Model hyperparameters and settings.}
{
\begin{tabular}{@{}ll@{}}
\toprule
\textbf{Hyperparameter} & \textbf{Value} \\ 
\midrule
Model Layers    & 2 \\
Hidden Size         & 512 \\
Attention Heads    & 4 \\
Activation Function & SiLU \\
Attention Dropout   & 0.1  \\
RMS Norm Eps    & 1e-6  \\
Codebook Num      & 32  \\
Codebook Size & 256 \\
Learning Rate       & 0.003 \\
Optimizer           & Adam \\
Semantic Encoder    & sentence-t5-base \\
Semantic Embedding Dimension & 768 \\
Batch Size          & 256 \\
Warmup Steps     & 1000 \\
Epoch        & 200 \\
Early Stopping Patience & 20 \\
Early Stopping Validate Metric & NDCG@20  \\
Beam Size          & 200 \\
Random Seed       & 2025 \\
$\alpha$     & 1.0 \\
$\beta$        & 0.01 \\
$\tau$       & 0.03 \\
\bottomrule
\end{tabular}
}
\label{tab:model_params}
\end{table}
Table \ref{tab:model_params} details the practical hyperparameter configurations for FAIR. All experiments were conducted on a single NVIDIA Tesla V100 GPU with 32GB memory, using implementations based on PyTorch 2.5.1 and the HuggingFace library\footnote{https://huggingface.co/}. For moderate-scale datasets such as Beauty, Sports, and Toys, each training run with the reported hyperparameters required less than 1 GPU hour, while a single training run on the large-scale CDs dataset took approximately 6 GPU hours. We carefully tuned several key hyperparameters on the moderate-scale datasets, including the learning rate, model layers, attention heads, hidden size, codebook number, codebook size and the loss coefficients $\alpha$ and $\beta$. Due to computational resource constraints, we did not perform exhaustive tuning over all adjustable hyperparameters; thus, the reported results may not reflect the absolute optimal performance of FAIR. However, since the remaining hyperparameters are not central to the core contributions of FAIR, we did not prioritize extensive tuning of them.

\section{Sensitivity of the Loss Coefficients $\alpha$ and $\beta$}
\label{Appendix:sensitivity}
\begin{figure}[htbp]
    \centering
    \includegraphics[width=\linewidth]{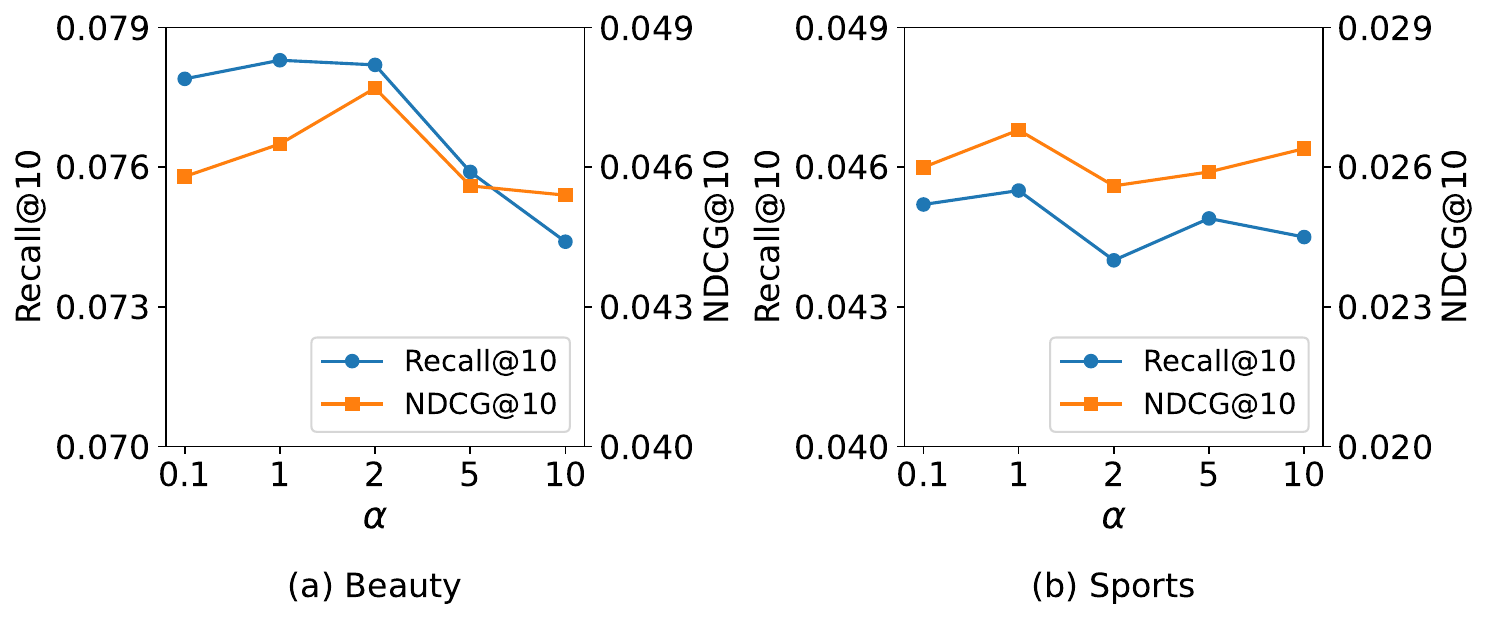}
    \caption{Sensitivity of the loss coefficient $\alpha$.}
    \label{fig:alpha}
\end{figure}

\begin{figure}[htbp]
    \centering
    \includegraphics[width=\linewidth]{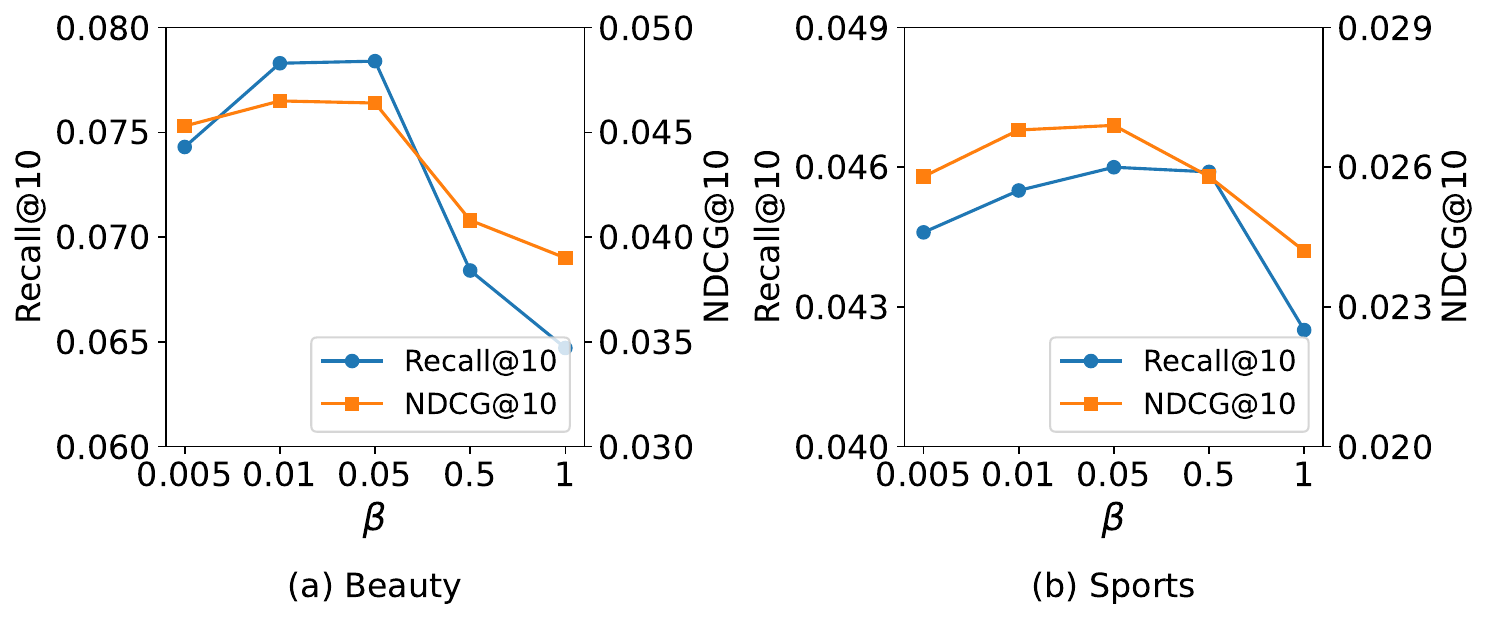}
    \caption{Sensitivity of the loss coefficient $\beta$.}
    \label{fig:beta}
\end{figure}

We perform a sensitivity analysis on the loss coefficients $\alpha$ and $\beta$ to assess their influence on model performance across diverse datasets. As illustrated in Figures \ref{fig:alpha} and \ref{fig:beta}, both Recall@10 and NDCG@10 demonstrate relative stability over a wide range of parameter values, suggesting that the proposed method is not highly sensitive to precise hyperparameter configurations. In particular, for $\alpha$, performance generally attains its optimum or remains competitive around $\alpha = 1$ on both Beauty and Sports datasets, with larger values leading to mild performance degradation. Regarding $\beta$, we find that moderate values consistently enhance performance, whereas excessively large values can adversely affect results, especially in terms of Recall@10. To streamline hyperparameter tuning and ensure cross-dataset consistency, we adopt a unified configuration of $\alpha = 1$ and $\beta = 0.01$ throughout all experiments. This selection provides a favorable balance between performance robustness and practical applicability, eliminating the need for dataset-specific tuning and underscoring the generalizability of our approach.

\section{Sensitivity of the Dropout Rate.}
\begin{figure}[htbp]
    \centering
    \includegraphics[width=\linewidth]{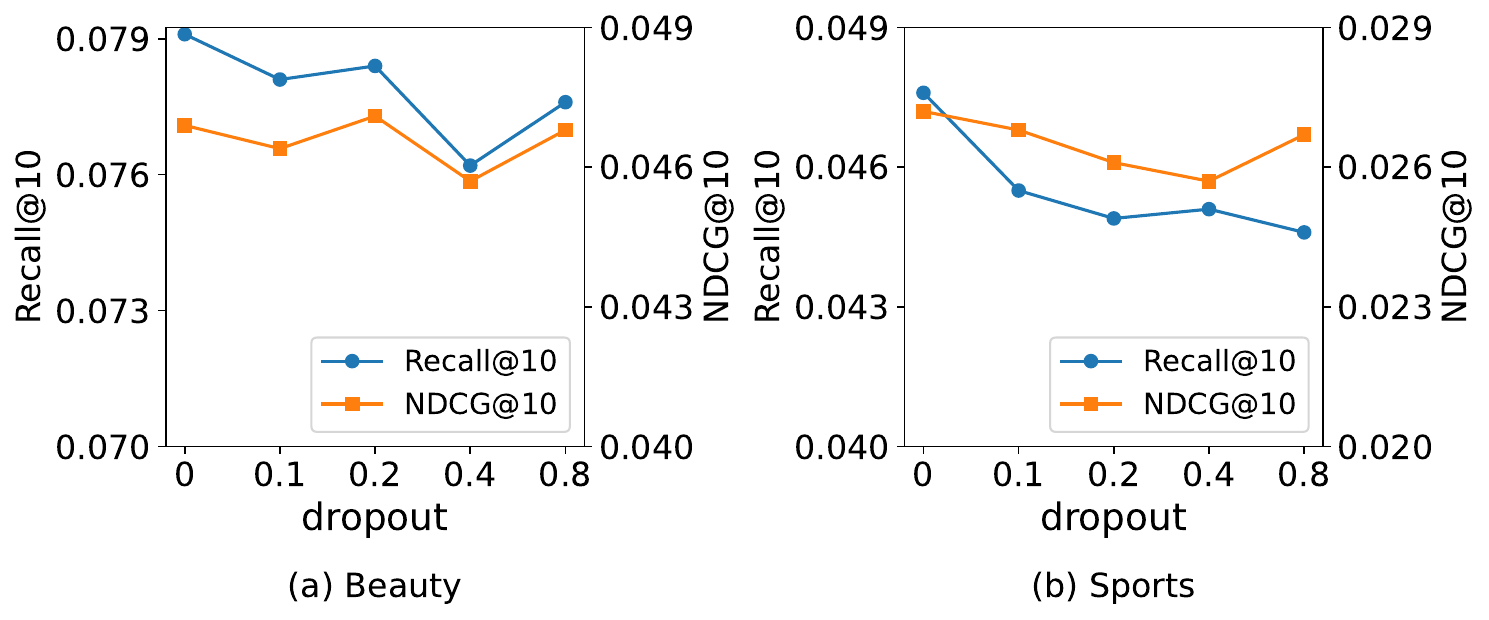}
    \caption{Sensitivity of the dropout rate.}
    \label{fig:dropout}
\end{figure}
To examine whether additional stochastic regularization offers complementary benefits to FAIR, we further analyze its sensitivity to the dropout rate. As shown in Figure \ref{fig:dropout}, on both the Beauty and Sports datasets, FAIR achieves the best performance in Recall@10 and NDCG@10 when the dropout rate is set to 0. As the dropout rate increases, performance gradually declines or shows only marginal fluctuation without improvement, a trend particularly evident on the Sports dataset. These observations indicate that FAIR is already able to learn sufficiently discriminative and robust representations. Introducing additional randomness via dropout may instead disrupt the learned attention allocation and weaken the model's ability to consistently focus on informative signals. Thus, dropout no longer serves as an effective regularizer but acts as an implicit source of noise. These results suggest that the explicit noise robustness modeling in FAIR can largely replace conventional stochastic regularization, highlighting that directly reinforcing noise resistance at the objective level is more effective than relying on generic regularization techniques. Since analyzing dropout sensitivity is not the primary focus of this work but rather provides diagnostic insight, we follow default settings and fix the dropout rate to 0.1 in all other experiments.

\section{Sensitivity of $p_{\text{mask}}$ and $p_{\text{sub}}$.}
\begin{figure}[htbp]
    \centering
    \includegraphics[width=\linewidth]{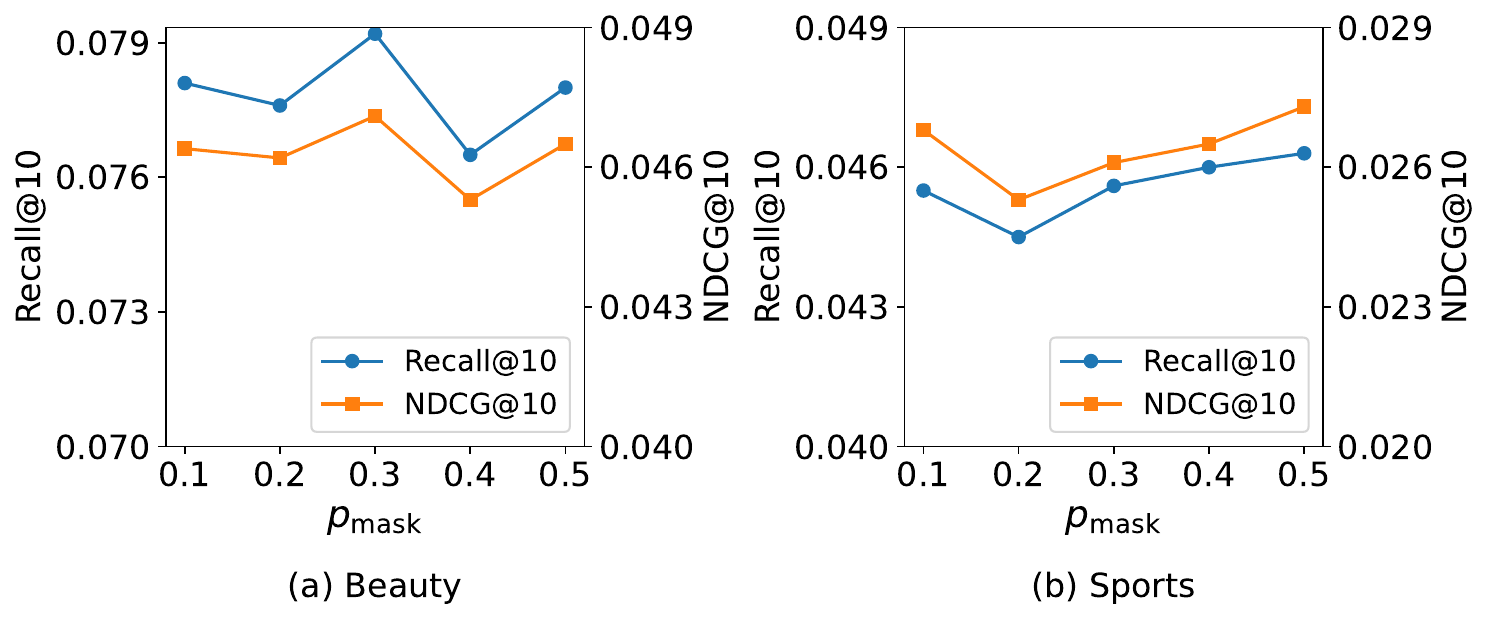}
    \caption{Sensitivity of $p_{\text{mask}}$.}
    \label{fig:p_mask}
\end{figure}

\begin{figure}[htbp]
    \centering
    \includegraphics[width=\linewidth]{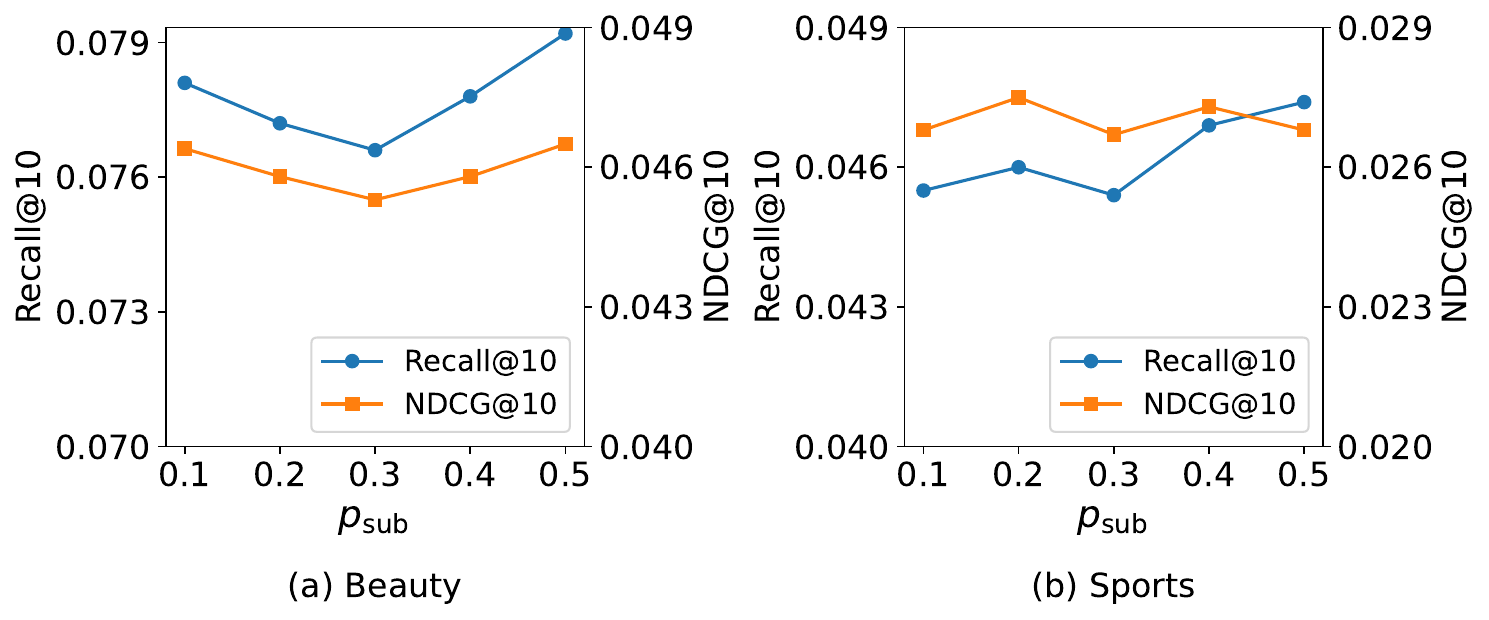}
    \caption{Sensitivity of $p_{\text{sub}}$.}
    \label{fig:p_sub}
\end{figure}
We investigate the impact of the masking probability $p_{\text{mask}}$ and substitution probability $p_{\text{sub}}$, varied from $0.1$ to $0.5$, on Beauty and Sports datasets, as shown in Figures \ref{fig:p_mask} and \ref{fig:p_sub}. For $p_{\text{mask}}$, performance on Beauty peaks at an intermediate value and slightly degrades with excessive masking, while on Sports both Recall@10 and NDCG@10 show a generally increasing trend, indicating that a moderate level of masking helps emphasize informative contexts. Regarding $p_{\text{sub}}$, performance on Beauty first decreases and then recovers as the probability grows, whereas on Sports both metrics vary mildly without a clear monotonic trend; overall, an appropriate degree of substitution proves beneficial, while overly small or large values can be suboptimal depending on the dataset. Since this analysis is intended for diagnostic insight rather than hyperparameter optimization, we fix $p_{\text{mask}} = p_{\text{sub}} = 0.1$ in all other experiments.

\section{Computational Costs Comparison}
\begin{table}[htbp]
    \caption{Comparison of model parameters and FLOPs.}
    \centering
    \begin{tabular}{lccc}
        \toprule
         & HSTU & RPG & FAIR \\ 
        \midrule
        Parameters (M) & 17 & 19 & 22 \\ 
        GFLOPs     & 234 & 242 & 282 \\ 
        \bottomrule
    \end{tabular}
    \label{tab:computation costs}
\end{table}

A comparative analysis of computational efficiency is conducted between FAIR and two representative decoder-based generative recommendation models, namely HSTU \cite{zhai2024actions} and RPG \cite{hou2025generating}, with respect to model size and computational complexity. As summarized in Table \ref{tab:computation costs}, FAIR comprises 22\,M parameters and requires 282\,GFLOPs, representing a moderate increase over HSTU (17\,M parameters, 234\,GFLOPs) and RPG (19\,M parameters, 242\,GFLOPs). This additional cost stems primarily from architectural distinctions between FAIR and their frameworks. Nevertheless, both the parameter count and computational overhead of FAIR remain within the same order of magnitude as those of the benchmarked models. It is noteworthy that, despite comparable model scale and FLOPs, FAIR consistently delivers substantially superior recommendation performance across all evaluated datasets. These results demonstrate that the performance gains achieved by FAIR are not attributable to a significant expansion in model parameters or computational burden, but rather to a more efficient utilization of the model’s capacity. Specifically, FAIR endows Transformer-based generative models, for the first time, with the ability to enhance attention to relevant context while suppressing noise, thereby enabling more effective representation learning.

\end{document}